\documentclass[11pt, oneside]{article}

\usepackage{amssymb}
\usepackage{amsmath}
\usepackage{amsthm}
\usepackage{graphicx}
\usepackage{fullpage}

\usepackage{algorithm}
\usepackage{algorithmic}

\def\Dmin{\lambda}
\def\Dmax{\Lambda}

\begin{document}
\newtheorem{problem}{Problem}
\newtheorem{definition}[problem]{Definition}
\newtheorem{proposition}[problem]{Proposition}
\newtheorem{theorem}[problem]{Theorem}
\newtheorem{corollary}[problem]{Corollary}
\newtheorem{remark}[problem]{Remark}
\newtheorem{lemma}[problem]{Lemma}
\newtheorem{conjecture}[problem]{Conjecture}

\title{Revisiting the Problem of Searching on a Line\thanks{This research has been partially funded by NSERC and FQRNT. A preliminary version appeared in~\cite{DBLP:conf/esa/BoseCD13}.}}

\author{Prosenjit Bose\thanks{School of Computer Science, Carleton University, Ottawa, Canada}
\and Jean-Lou De Carufel$^\dagger$
\and Stephane Durocher\thanks{Department of Computer Science, University of Manitoba, Winnipeg, Canada}}

\maketitle

\begin{abstract}
We revisit the problem of searching for a target at an unknown location on a 
line when given upper and lower bounds on the distance $D$
that separates the initial position of the searcher from the target.
Prior to this work,
only asymptotic bounds were known for the optimal competitive ratio 
achievable by any search strategy in the worst case.
We present the first tight bounds on the exact optimal competitive ratio 
achievable, parameterized in terms of the given bounds on $D$,
along with an optimal search strategy that achieves this competitive ratio.
We prove that this optimal strategy is unique. We characterize the conditions under which an 
optimal strategy can be computed exactly and, when it cannot, 
we explain how numerical methods can be used efficiently.
In addition, we answer several related open questions,
including the \emph{maximal reach problem},
and we discuss how to generalize these results to $m$ rays, 
for any $m \geq 2$.
\end{abstract}

\section{Introduction}
\label{section introduction}

Search problems are broadly studied within computer science. 
A fundamental search problem, which is the focus of this paper,
is to specify how a searcher should move
to find an immobile target at an unknown location on a line such that the 
total relative distance travelled by the searcher is minimized in the worst case
\cite{DBLP:journals/iandc/Baeza-YatesCR93,%
DBLP:journals/dam/HipkeIKL99,%
DBLP:journals/tcs/Lopez-OrtizS01}.
The searcher is required to move continuously on the line,
i.e., discontinuous jumps, such as random access in an array, are not possible.
Thus, a search corresponds to a sequence of alternating
left and right displacements by the searcher.
This class of geometric search problems 
was introduced by Bellman~\cite{bellman56}
who first formulated the problem of searching for the boundary of a region from
an unknown random point within its interior.
Since then, many variants of the line search problem have been studied, 
including multiple rays sharing a common endpoint
(as opposed to a line, which corresponds to two rays),
multiple targets, multiple searchers, 
moving targets,
and randomized search strategies
(e.g., \cite{alpern1999,%
alpern2003theory,%
alpern2013book,%
DBLP:journals/iandc/Baeza-YatesCR93,%
DBLP:conf/stoc/BenderFRSV98,%
DBLP:conf/icalp/CollinsCGL10,%
DBLP:journals/tcs/CzyzowiczILP11,%
DieudonneP13,%
DBLP:journals/comgeo/HammarNS01,%
DBLP:conf/icalp/KoutsoupiasPY96,%
DBLP:journals/tcs/Lopez-OrtizS01,%
DBLP:journals/tcs/MarcoGKKPV06}).

For any given search strategy $f$ and any given target location, 
we consider the ratio $A/D$, 
where $A$ denotes the total length of the search path travelled by a searcher
before reaching the target by applying strategy $f$,
and $D$ corresponds to the minimum travel distance necessary
to reach the target.
That is, the searcher and target initially lie a distance $D$ from each other
on a line,
but the searcher knows neither the value $D$ nor whether the target lies 
to its left or right.
The {\em competitive ratio}
of a search strategy $f$, denoted $CR(f)$, is measured by the supremum 
of the ratios achieved over all possible target locations.
Observe that $CR(f)$ is unbounded
if $D$ can be assigned any arbitrary real value;
specifically, the searcher must know a lower bound $\Dmin \leq D$.
Thus, it is natural to consider scenarios where the searcher has
additional information about the distance to the target.
In particular, in many instances the searcher can estimate 
good lower and upper bounds on $D$.
Given a lower bound $D \geq \Dmin$,
Baeza-Yates et al.~\cite{DBLP:journals/iandc/Baeza-YatesCR93}
show that any optimal strategy achieves 
a competitive ratio of $9$.
They describe such a strategy, which we call the {\em Power of Two} strategy.
Furthermore, they observe that when $D$ is known to the searcher,
it suffices to travel a distance of $3D$ in the worst case, achieving a 
competitive ratio of $3$.

We represent a search strategy 
by a function $f:\mathbb{N} \rightarrow \mathbb{R}^+$.
Given such a function,
a searcher travels a distance of $f(0)$ in one direction from the origin 
(say, to the right),
returns to the origin,
travels a distance of $f(1)$ in the opposite direction (to the left), 
returns to the origin, and so on, until reaching the target.
We refer to $f(i)$ as the distance the searcher travels
from the origin during the $i$-th iteration.
The corresponding function for the Power of Two strategy of Baeza-Yates et al.\
is $f(i) = 2^i \Dmin$.
Showing that every optimal strategy achieves a competitive ratio of exactly 9 
relies on the fact that no upper bound on $D$ is specified
\cite{DBLP:journals/iandc/Baeza-YatesCR93}.
Therefore, it is natural to ask whether a search strategy can achieve
a better competitive ratio when provided lower and upper bounds 
$\Dmin \leq D \leq \Dmax$.

Given $R$, the \emph{maximal reach problem}
examined by Hipke et al.~\cite{DBLP:journals/dam/HipkeIKL99}
is to identify the largest bound $\Dmax$ such that there exists a search strategy
that finds any target within distance $D \leq \Dmax$
with competitive ratio at most $R$.
L\'opez-Ortiz and Schuierer~\cite{DBLP:journals/tcs/Lopez-OrtizS01}
study the maximal reach problem on $m$ rays,
from which they deduce that the competitive ratio $CR(f_{opt})$ of any optimal strategy $f_{opt}$ is at least
$$1+2 \frac{m^m}{(m-1)^{m-1}}-O\left(\frac{1}{\log^2\rho}\right) ,$$
where $\rho = \Dmax/\Dmin$.
When $m = 2$, the corresponding lower bound becomes
$$9-O\left(\frac{1}{\log^2\rho}\right) .$$
They also provide a general strategy that achieves 
this asymptotic behaviour on $m$ concurrent rays, given by
$$f(i) = \sqrt{1+\frac{i}{m}}\,\left(\frac{m}{m-1}\right)^i\Dmin .$$
Again,
for $m=2$ this is
$$f(i) = \sqrt{1+\frac{i}{2}}\,2^i\Dmin .$$
Surprisingly, this general strategy is independent of $\rho$.
In essence, it ignores any upper bound on $D$, 
regardless of how tight it is.
Thus, we examine whether there exists a better search strategy 
that depends on $\rho$, thereby using both the upper and lower bounds on $D$.
Furthermore, previous lower bounds on $CR(f_{opt})$ have 
an asymptotic dependence on $\rho$ applying only to large values of $\rho$,
corresponding to having only coarse bounds on $D$.
Can we express tight bounds on $CR(f_{opt})$ in terms of $\rho$?

Let $f_{opt}(i) = a_i\Dmin$ denote an optimal strategy for given values $\Dmin$ and $\Dmax$.
Since $f_{opt}$ is optimal and $D \geq \Dmin$,
we must have $f_{opt}(i) \geq \Dmin$ for all $i \geq 0$.
Therefore,
$a_i \geq 1$ for all $i \geq 0$.
Moreover,
for any possible position of the target,
the strategy $f_{opt}$ must eventually reach it.
Hence,
there must be two integers $i$ and $i'$ of different parities such that
$f_{opt}(i) \geq \Dmax$ and $f_{opt}(i') \geq \Dmax$.
Equivalently,
there must be two integers $i$ and $i'$ of different parities such that
$a_i \geq \rho$ and $a_{i'} \geq \rho$.
Since $f_{opt}$ is optimal,
we have $f_{opt}(i) = f_{opt}(i') = \Dmax$.
Moreover,
let $n$ be the smallest integer such that $f_{opt}(n) = \Dmax$
(equivalently, $a_n = \rho$).
Since $f_{opt}$ is optimal,
we have $f_{opt}(n+1) = \Dmax$
(equivalently, $a_{n+1} = \rho$).
Consequently,
$n+2$ is the number of iterations necessary to reach the target with strategy $f_{opt}$ in the worst case
(recall that the sequence starts at $i = 0$).
The question is now to determine the sequence $\{a_i\}_{i=0}^n$
that defines $f_{opt}$.
L\'opez-Ortiz and Schuierer~\cite{DBLP:journals/tcs/Lopez-OrtizS01}
provide an algorithm to compute the maximal reach for a given competitive ratio
together with a strategy corresponding to this maximal reach.
They state that the value $n$ and the sequence $\{a_i\}_{i=0}^n$ 
can be computed using binary search,
which increases the running time proportionally to $\log\rho$.
Can we find a faster algorithm for computing $f_{opt}$?
Since in general,
$a_0$ is the root of a polynomial equation of unbounded degree 
(see Theorem~\ref{theorem strategy with n cuts}), 
a binary search is equivalent to the bisection method
for solving polynomial equations.
However,
the bisection method is a slowly converging numerical method.
Can the computational efficiency be improved?
Moreover, given $\varepsilon$, 
can we bound the number of steps necessary for a root-finding algorithm 
to identify a solution within tolerance $\varepsilon$ of the exact value?

\subsection{Overview of Results}
We address all of the questions raised above. 
We characterize $f_{opt}$
by computing the sequence $\{a_i\}_{i=0}^n$ for the optimal strategy.
We do this by computing the number of iterations $n+2$ needed to find the target in the worst case.
We can compute $n$ in $O(1)$ time since
we prove that 
$n \in \{ \lfloor \log_2\rho\rfloor -1, \lfloor \log_2\rho\rfloor \}$,
where $\rho =  \Dmax/\Dmin $.
Then,
we define a family of $n+1$ polynomials $p_0, \ldots, p_n$, 
where $p_i$ has degree $i+1$.
We show that $a_0$ is the largest real solution
to the polynomial equation $p_n(x)=\rho$.
Each of the remaining elements in the sequence $\{a_i\}_{i=0}^n$
can be computed in $O(1)$ time since we prove that $a_1 = a_0(a_0-1)$
and $a_i = a_0(a_{i-1}-a_{i-2})$ for $2\leq i \leq n$.
This also shows that the optimal strategy is unique.
However,
as we show in Proposition~\ref{proposition infinite family m > 2},
when no upper bound is known there exist infinitely many optimal strategies
for any $m \geq 2$.

We give an exact characterization of $f_{opt}$ and show that
$CR(f_{opt}) = 2a_0 +1$. 
This allows us to establish the following 
bounds on the competitive ratio of an optimal strategy in terms of $\rho$:
$$8\cos^2\!\left(\frac{\pi}{\lceil\log_2\rho\rceil+2}\right)+1 \leq CR(f_{opt}) < 8\cos^2\!\left(\frac{\pi}{\lfloor\log_2\rho\rfloor+4}\right)+1 .$$

L\'opez-Ortiz and Schuierer~\cite{DBLP:journals/tcs/Lopez-OrtizS01}
show that $CR(f_{opt}) \to 9$ as $\rho \to \infty$.
We show that $f_{opt} \to f_{\infty}$ as $\rho \rightarrow \infty$,
where $f_{\infty}(i) = (2i+4)2^i \Dmin$ has a competitive ratio of $9$.
We thereby obtain an alternate proof of the result of 
Baeza-Yates et al.~\cite{DBLP:journals/iandc/Baeza-YatesCR93}.
The strategy $f_{\infty}$ is a member
of the infinite family of optimal strategies in the unbounded case
which we describe in Proposition~\ref{proposition infinite family m > 2}.

We assume the Real RAM model of computation, including $k$-th roots, 
logarithms, exponentiation, and trigonometric functions \cite{DBLP:books/sp/PreparataS85}.
The computation of each term $a_i$ in the sequence defining $f_{opt}$
involves computing the largest real root of a polynomial equation 
of degree $n+1$.
We prove that
$n+1 \leq 4$
if and only if $\rho \leq 32\cos^5(\pi/7) \approx 18.99761$.
In this case the root can be expressed exactly using only the operations
$+$, $-$, $\times$, $\div$, $\sqrt{\cdot}$ and $\sqrt[3]{\cdot}$.
This implies that if $\Dmax \leq 32\cos^5(\pi/7)\Dmin$,
then $f_{opt}$ can be computed exactly in $O(1)$ time
($O(1)$ time per $a_i$ for $0\leq i \leq n < 4$).
In general, when $n+1 \geq 5$, Galois theory implies that 
the equation $p_n(x) = \rho$ cannot be solved by radicals.
Since the corresponding polynomials have unbounded degree,
we are required to consider approximate solutions when $\rho > 32\cos^5(\pi/7)$.
Therefore, we explain how to find a solution $f_{opt}^*$,
such that $CR(f_{opt}^*) \leq CR(f_{opt}) + \varepsilon$ 
for a given tolerance $\varepsilon$.

If $n \geq 7\varepsilon^{-1/3}-4$,
we give an explicit formula for $a_0$.
Hence,
an {$\varepsilon$-approxima\-tion} can be computed in $O(n)=O(\log \rho)$ time
($O(1)$ time per $a_i$ for $0\leq i \leq n$).
Otherwise,
if $32\cos^5(\pi/7) < n < 7\varepsilon^{-1/3}-4$,
we show that $a_0$ lies in an interval of length at most $7^3\,(n+4)^{-3}$.
Moreover,
we prove that the polynomial is strictly increasing on this interval.
Hence, usual root-finding algorithms work well.
Given $a_0$,
the remaining elements of the sequence $\{a_1, \ldots, a_n\}$ 
can be computed in $O(n)$ time ($O(1)$ time per $a_i$ for $1\leq i \leq n$).
This is all summarized in Algorithm~\ref{algorithm optimal strategy}.
\begin{algorithm}
\caption{Optimal Strategy for Searching on a Line}
\label{algorithm optimal strategy}
\begin{algorithmic}[1]
\REQUIRE An interval $[\Dmin,\Dmax]$ that contains $D$ and a tolerance $\varepsilon > 0$.
\ENSURE A sequence $\{a_i\}$ defining a search
strategy $f$ and its corresponding competitive ratio,
where $f$ is an exact optimal strategy
if $\rho= \Dmax/\Dmin  \leq 32\cos^5(\pi/7) \approx 18.99761$,
and $f$ has competitive ratio at most $\varepsilon$ above optimal otherwise.
	\STATE $\rho = \Dmax/\Dmin$; $n=\lfloor \log_2 \rho \rfloor$; $\gamma = 2\cos\left(\frac{\pi}{n+3}\right)$; $a_{-1} = 1$
	\IF{$n+1 > \log_{\gamma}\rho$}
		\STATE $n = n-1$
	\ENDIF
	\STATE  $\alpha_{n+1}=4\cos^2 \left(\frac{\pi}{n+3}\right)$; $\alpha_{n+2}=4\cos^2 \left(\frac{\pi}{n+4}\right)$
	\IF{$n \leq 3$}
		\STATE /* Find an exact solution for $a_0$ in $O(1)$ time. */
		\STATE Find $a_0\in\left[\alpha_{n+1},\alpha_{n+2}\right]$ such that $p_n(a_0) = \rho$.
	\ELSIF{$n \geq 7\varepsilon^{-1/3}-4$}
		\STATE /* Find an approximate solution for $a_0$ in $O(1)$ time. */
		\STATE $a_0 = \alpha_{n+2}$
	\ELSE
		\STATE /* Find an approximate solution for $a_0$ using numerical methods. */
		\STATE Find $a_0\in\left[\alpha_{n+1},\alpha_{n+2}\right]$ such that $\left|p_n(a_0) - \rho\right| < \varepsilon$, using numerical methods.
	\ENDIF
	\FOR{$1 \leq i \leq n$}
		\STATE $a_i = a_0(a_{i-1}-a_{i-2})$
	\ENDFOR
	\RETURN $\{a_0\Dmin,a_1\Dmin,...,a_n\Dmin,\Dmax\}$ and $2a_0+1$
\end{algorithmic}
\end{algorithm}
The results presented in Section~\ref{section searching bounded line}
establish the correctness of Algorithm~\ref{algorithm optimal strategy}.

In Section~\ref{section maximal reach},
we explain how to solve the maximal reach problem using the results of Section~\ref{section searching bounded line}.
Finally,
we explain how our technique can be generalized to $m$ rays
in Section~\ref{section searching m rays}.

\section{Searching on a Bounded Line}
\label{section searching bounded line}

As we explained in Section~\ref{section introduction},
we are looking for a sequence of numbers $\{a_i\}_{i=0}^n$ that defines an optimal strategy
(recall that we have $a_n = \rho$)
$$f_{opt}(i) = 
\begin{cases} a_i\Dmin & \text{if $0 \leq i < n$,} \\
\rho\Dmin & \text{if $i \geq n$.}
\end{cases}$$
L\'opez-Ortiz and Schuierer~\cite{DBLP:journals/tcs/Lopez-OrtizS01}
showed that there always exists an optimal strategy for $m$ rays ($m\geq 2$)
that is \emph{periodic} and \emph{monotone}.
Let the rays be labelled from $0$ to $m-1$.
A strategy is periodic if after visiting the ray $k$
(for any $0 \leq k \leq m-1$),
it visits the ray $(k+1) \pmod{m}$.
A strategy is monotone if the values in the sequence $\{a_i\}_{i=0}^n$ are non-decreasing:
$a_i \leq a_{i+1}$ for all $0 \leq i \leq n-1$.

Let
$$\phi(f,D) = 2\sum_{i=0}^{f^{-1}(D)} f(i) + D$$
denote the cost incurred by a strategy $f$ to find a target at distance $D$ in
the worst case
and $f^{-1}(D)$ be the smallest integer $j$ such that $f(j) \geq D$.
Our goal is to identify a sequence of numbers $\{a_i\}_{i=0}^n$ that defines 
a periodic and monotone strategy $f_{opt}$ which minimizes
$$CR(f_{opt}) = \sup_{D \in [\Dmin,\Dmax]} \frac{\phi(f_{opt},D)}{D} .$$
At first,
for each $n \geq 0$,
we find the optimal strategy $f_n$ that takes $n+2$ iterations in the worst case
for a given $\rho$
(refer to Theorem~\ref{theorem strategy with n cuts}).
Then,
we explain how to compute the optimal number of iterations $n_{opt}+2$ for a given $\rho$
(refer to Theorem~\ref{theorem optimal n}),
from which $f_{opt} = f_{n_{opt}}$.
We first focus on the cases $n=0$, $n=1$ and $n=2$.
Then we characterize the optimal strategy $f_n$ for a general $n$.

If $n=0$,
then we must have $f_0(i)=\Dmax$.
The competitive ratio of $f_0$ is then
$$CR(f_0) = \sup_{D \in [\Dmin,\Dmax]} \frac{\phi(f_0,D)}{D} = \sup_{D \in [\Dmin,\Dmax]}\frac{2\Dmax+D}{D} = 2\rho+1  .$$
Observe that $f_0$ is optimal when $\rho = 1$,
i.e.,
when $D$ is known.
When $n\geq 1$,
finding a sequence of numbers $\{a_i\}_{i=0}^n$
corresponds to partitioning the interval $[\Dmin,\Dmax]$
into $n+1$ subintervals.
At each iteration,
the corresponding subinterval
represents a set of candidate locations for the target.

If $n=1$,
this corresponds to cutting $[\Dmin,\Dmax]$ once
at a point $\Dmin \leq a_0 \Dmin \leq \rho \Dmin = \Dmax$.
Namely,
we search a sequence of two intervals, 
$[\Dmin,a_0\Dmin]$ and $(a_0\Dmin,\rho\Dmin] = (a_0\Dmin, \Dmax]$,
from which we define
$$f_1(i) = 
\begin{cases} a_0\Dmin & \text{if $0 \leq i < 1$,} \\ 
\rho\Dmin & \text{if $i \geq 1$.} 
\end{cases}$$
Therefore,
$a_0$ needs to be chosen such that $CR(f_1)$ is minimized.
We have
\begin{align*}
\sup_{D \in [\Dmin,a_0\Dmin]} \frac{\phi(f_1,D)}{D} &= \sup_{D \in [\Dmin,a_0\Dmin]} \frac{2a_0\Dmin + D}{D} = 2a_0+1 ,\\
\sup_{D \in (a_0\Dmin,\rho\Dmin]} \frac{\phi(f_1,D)}{D} &= \sup_{D \in [a_0\Dmin,\rho\Dmin]} \frac{2a_0\Dmin + 2\rho\Dmin + D}{D} = 3 + 2\frac{\rho}{a_0} .
\end{align*}
Hence,
to minimize $CR(f_1)$,
we must select $a_0$,
where $1 \leq a_0 \leq \rho$,
such that
$$2a_0+1 = 3 + 2\frac{\rho}{a_0} .$$
Therefore, $a_0 = (1+\sqrt{1+4\rho})/2$ and $CR(f_1) = 2+\sqrt{1+4\rho}$.
We have that $CR(f_0) \leq CR(f_1)$ if and only if $1 \leq \rho \leq 2$.

If $n=2$,
this corresponds to cutting $[\Dmin,\Dmax]$ twice
at points $\Dmin \leq a_0 \Dmin \leq a_1\Dmin \leq \rho \Dmin = \Dmax$.
Namely,
we search a sequence of three intervals, 
$[\Dmin,a_0\Dmin]$, $(a_0\Dmin,a_1\Dmin]$ and $(a_1\Dmin, \Dmax]$,
from which we define
$$f_2(i) = 
\begin{cases} a_0\Dmin & \text{if $0 \leq i < 2$,} \\ 
\rho\Dmin & \text{if $i \geq 2$.} 
\end{cases}$$
Therefore,
$a_0$ and $a_1$ need to be chosen such that $CR(f_2)$ is minimized.
We have
\begin{align*}
\sup_{D \in [\Dmin,a_0\Dmin]} \frac{\phi(f_2,D)}{D} &= \sup_{D \in [\Dmin,a_0\Dmin]} \frac{2a_0\Dmin + D}{D} = 2a_0+1 ,\\
\sup_{D \in [a_0\Dmin,a_1\Dmin]} \frac{\phi(f_2,D)}{D} &= \sup_{D \in [a_0\Dmin,a_1\Dmin]} \frac{2a_0\Dmin + 2a_1\Dmin + D}{D} = 3 + 2 \frac{a_1}{a_0} ,\\
\sup_{D \in (a_1\Dmin,\rho\Dmin]} \frac{\phi(f_2,D)}{D} &= \sup_{D \in [a_1\Dmin,\rho\Dmin]} \frac{2a_0\Dmin + 2a_1\Dmin + 2\rho\Dmin + D}{D} = 3 + 2\frac{a_0}{a_1} + 2\frac{\rho}{a_1} .
\end{align*}
Hence,
to minimize $CR(f_2)$,
we must select $a_0$ and $a_1$,
where $1 \leq a_0 \leq a_1 \leq \rho$,
such that
\begin{align*}
2a_0+1 &= 3 + 2 \frac{a_1}{a_0} ,\\
2a_0+1 &= 3 + 2\frac{a_0}{a_1} + 2\frac{\rho}{a_1} .
\end{align*}
Therefore,
$a_0^3 - 2a_0^2 = \rho$,
from which
\begin{align*}
a_0 &= \frac{1}{3} \left(2+\frac{4}{\left(8+\frac{27 \rho}{2}+\frac{3}{2} \sqrt{3} \sqrt{\rho (32+27 \rho)}\right)^{1/3}}+\left(8+\frac{27\rho}{2}+\frac{3}{2} \sqrt{3} \sqrt{\rho (32+27 \rho)}\right)^{1/3}\right) ,\\
a_1 &= a_0(a_0-1) ,\\
CR(f_2) &= 2a_0 + 1 .
\end{align*}
We have that $CR(f_1) \leq CR(f_2)$ if and only if $1 \leq \rho \leq 2+\sqrt{5}$.

In general,
we can partition the interval $[\Dmin,\rho\Dmin]$ into $n+1$ subintervals
whose endpoints correspond to the sequence
$\Dmin, a_0 \Dmin, \ldots, a_{n-1}\Dmin,\rho\Dmin$,
from which we define
$$f_n(i) = \begin{cases} a_i\Dmin & \text{if $0\leq i < n$,} \\ 
\rho\Dmin & \text{if $i \geq n$.} \end{cases}$$
Therefore,
we must select $a_0, \ldots, a_{n-1}$, where
$1 \leq a_0 \leq a_1 \leq \ldots \leq a_{n-1} \leq \rho$,
such that $CR(f_n)$ is minimized.
We have
\begin{align*}
\sup_{D \in [\Dmin,a_0\Dmin]} \frac{\phi(f_n,D)}{D} &= 2a_0+1  , \\
\sup_{D \in (a_i\Dmin,a_{i+1}\Dmin]} \frac{\phi(f_n,D)}{D} &= 1 + 2\sum_{k=0}^{i+1} \frac{a_k}{a_i} \quad (0 \leq i \leq n-2),\\
\sup_{D \in (a_{n-1}\Dmin,\rho\Dmin]} \frac{\phi(f_n,D)}{D} &= 1 + 2\sum_{k=0}^{n-1} \frac{a_k}{a_{n-1}}+2\frac{\rho}{a_{n-1}}  .
\end{align*}
Hence,
the values $a_i$
are solutions to the following system of equations:
\begin{align}
\label{eqn basics 1}
1 + 2\sum_{k=0}^{i+1} \frac{a_k}{a_i} &= 2a_0+1 \quad (0 \leq i \leq n-2), \\
\label{eqn basics 2}
1 + 2\sum_{k=0}^{n-1} \frac{a_k}{a_{n-1}}+2\frac{\rho}{a_{n-1}} &= 2a_0+1 . 
\end{align}
In Theorem~\ref{theorem strategy with n cuts},
we explain how to calculate the values $a_i$.
We prove that the solution to this system of equations can be obtained using
the following family of polynomials:
\begin{align}
\nonumber
p_0(x) &= x  ,\\
\nonumber
p_1(x) &= x(x - 1)  ,\\
\label{eqn def pi}
p_i(x) &= x\left(p_{i-1}(x) - p_{i-2}(x)\right) \qquad (i \geq 2).
\end{align}
We apply~\eqref{eqn def pi} without explicitly referring to it
when we manipulate the polynomials $p_i$.
Let $\alpha_i$ denote the largest real root of $p_i$ for each $i$.
\begin{theorem}
\label{theorem strategy with n cuts}
The following statement is true for all $n \in \mathbb{N}$.
\begin{enumerate}
\item\label{theorem strategy with n cuts item ai}
For all $0\leq i < n$,
the values $a_i$ that define $f_n$ satisfy $a_i=p_i(a_0)$.

\item\label{theorem strategy with n cuts item a0}
The number $a_0$ is the unique solution to the equation $p_n(x) = \rho$
such that $a_0 > \alpha_n$.

\item\label{theorem strategy with n cuts item an}
$a_n = p_n(a_0) = \rho$.

\item\label{theorem strategy with n cuts item cost}
$CR(f_n) = 2a_0+1$.
\end{enumerate}
\end{theorem}

To prove Theorem~\ref{theorem strategy with n cuts},
we need the following proposition and lemma.
\begin{proposition}
The following equalities are true for all $n \in \mathbb{N}$.
\begin{align}
\label{eqn pn}
p_{n+1}(x) &= xp_n(x) - \sum_{i=0}^np_i(x) \\
\label{factorization pn}
p_n(x) &= x^{\lfloor (n+1)/2\rfloor}\prod_{k=1}^{\lfloor (n+2)/2\rfloor}\left(x - 4\cos^2\!\left(\frac{k\pi}{n+2}\right)\right) \\
\label{formula alpha n}
\alpha_n &= 4\cos^2\!\left(\frac{\pi}{n+2}\right)
\end{align}
\end{proposition}

\proof
We prove~\eqref{eqn pn} by induction on $n$.
If $n = 0$,
\begin{align*}
xp_0(x) - \sum_{i=0}^0p_i(x) &= xp_0(x) - p_0(x) \\
&= x^2 - x \\
&= p_1(x)  .
\end{align*}

Suppose that the proposition is true for $n = \ell - 1$,
we now prove it for $n = \ell$.
\begin{align*}
& \phantom{=}
xp_{\ell}(x) - \sum_{i=0}^{\ell}p_i(x) \\
&= xp_{\ell}(x) - p_{\ell}(x) - \sum_{i=0}^{\ell-1}p_i(x) \\
&= xp_{\ell}(x) - p_{\ell}(x) + (p_{\ell}(x) - xp_{\ell-1}(x)) &&
\text{by the induction hypothesis,}\\
&= x(p_{\ell}(x) - p_{\ell-1}(x)) \\
&= p_{\ell+1}(x)
\end{align*}

Equation~\eqref{factorization pn} is a direct consequence of
Corollary 10 in~\cite{Hoggatt1974} since
the $p_n$'s are \emph{generalized Fibonacci polynomials}
(refer to~\cite{Hoggatt1974}).

Since by definition,
$\alpha_n$ is the largest real root of $p_n$,
\eqref{formula alpha n} is a direct consequence of~\eqref{factorization pn}.
\qed

\begin{lemma}
\label{lemma properties pn item sign of pn}
The following statement is true for all $n \in \mathbb{N}$.
For all $t \in \mathbb{R}$ such that $0 < t < \alpha_n$ and $p_n(t) > 0$,
there exists an $i\in\mathbb{N}$ such that $0 \leq i < n$ and $p_i(t) < 0$.
\end{lemma}

\proof
We consider six cases: (1) $n = 0$,
(2) $n = 1$,
(3) $n = 2$,
(4) $n = 3$,
(5) $n = 4$
and (6) $n \geq 5$.
\begin{enumerate}
\item Since $\alpha_0 = 0$ by~\eqref{formula alpha n},
then there does not exist a $t$ such that $0 < t < \alpha_0$.
Hence,
the statement is vacuously true.

\item Since $\alpha_1 = 1$ by~\eqref{formula alpha n}
and $p_1(x) < 0$ for any $0 < x < 1$,
then there does not exist a $t$ such that $0 < t < \alpha_1$
and $p_1(t) > 0$.
Hence,
the statement is vacuously true.

\item Since $\alpha_2 = 2$ by~\eqref{formula alpha n}
and $p_2(x) < 0$ for any $0 < x < 2$,
then there does not exist a $t$ such that $0 < t < \alpha_2$
and $p_2(t) > 0$.
Hence,
the statement is vacuously true.

\item By~\eqref{factorization pn},
$p_3(x) = x^2\left(x-(3-\sqrt{5})/2\right)\left(x-(3+\sqrt{5})/2\right)$.
Hence,
$t$ is such that $0 < t < \frac{3-\sqrt{5}}{2} < 1$.
Hence,
$p_1(t) = t(t - 1)< 0$
so that we can take $i=1$.

\item By~\eqref{factorization pn},
$p_4(x) = x^3(x-1)(x-3)$.
Hence,
$t$ is such that $0 < t < 1$.
Hence,
$p_1(t) = t(t - 1)< 0$
so that we can take $i=1$.

\item From~\eqref{factorization pn},
$t$ is such that
$$4\cos^2\!\left(\frac{(2\ell+1)\pi}{n+2}\right) < t < 4\cos^2\!\left(\frac{2\ell\pi}{n+2}\right) ,$$
for an $\ell$ satisfying $1 \leq \ell \leq \lfloor n/2\rfloor/2$.
We prove that the $i$ we need to pick is any integer in the interval
$$I = \left(\frac{n+2 - 4\ell}{2\ell},\frac{2n+2-4\ell}{2\ell+1}\right)  .$$
Notice that,
from elementary calculus,
we have
$$\min_{\stackrel{n \geq 5}{1 \leq \ell \leq \frac{1}{2} \lfloor\frac{n}{2}\rfloor}} \left(\frac{2n+2-4\ell}{2\ell+1} - \frac{n+2 - 4\ell}{2\ell}\right) = \frac{7}{6}  .$$
Hence,
if $n\geq 5$ and $1 \leq \ell \leq \lfloor n/2\rfloor/2$,
then there always exists an integer in $I$.
Also notice that for any integer $i\in I$,
then $0 \leq i < n$.
Hence,
for any $i\in I$,
\begin{align*}
\frac{n+2 - 4\ell}{2\ell} < &\;i < \frac{2n+2-4\ell}{2\ell+1}  ,\\
\frac{n-4\ell}{2\ell+1} < \frac{n+2 - 4\ell}{2\ell} < &\;i < \frac{2n+2-4\ell}{2\ell+1} < \frac{2n+4-4\ell}{2\ell}  ,\\
\frac{n-4\ell}{2\ell+1} < i < \frac{2n+2-4\ell}{2\ell+1} \quad&\text{and}\quad \frac{n+2 - 4\ell}{2\ell} < i < \frac{2n+4-4\ell}{2\ell}  ,\\
\frac{\pi}{i+2} < \frac{(2\ell+1)\pi}{n+2} < \frac{2\pi}{i+2} \quad&\text{and}\quad\frac{\pi}{i+2} < \frac{2\ell\pi}{n+2} < \frac{2\pi}{i+2}  ,
\end{align*}
from which
$$\frac{\pi}{i+2} < \frac{2\ell\pi}{n+2} < \frac{(2\ell+1)\pi}{n+2} < \frac{2\pi}{i+2}  ,$$
and hence,
$$4\cos^2\!\left(\frac{2\pi}{i+2}\right) < 4\cos^2\!\left(\frac{(2\ell+1)\pi}{n+2}\right) < t < 4\cos^2\!\left(\frac{2\ell\pi}{n+2}\right) < 4\cos^2\!\left(\frac{\pi}{i+2}\right) .$$
Consequently,
$p_i(t) < 0$ by~\eqref{factorization pn}.
\qed
\end{enumerate}

We now prove Theorem~\ref{theorem strategy with n cuts}.
\proof(Theorem~\ref{theorem strategy with n cuts})\
\begin{enumerate}
\item We prove this theorem by induction on $i$.
If $i=0$,
then $p_0(a_0) = a_0$.

Suppose the statement is true for any $i$ such that $0 < i < \ell < n$,
we now prove it for $i=\ell$.
From~\eqref{eqn basics 1},
we have
$$1 + 2\sum_{k=0}^{\ell}\frac{a_k}{a_{\ell-1}} = 2a_0+1 .$$
Therefore,
\begin{align*}
a_{\ell} &= a_0a_{\ell-1} - \sum_{k=0}^{\ell-1}a_k \\
 &= a_0 p_{\ell-1}(a_0) - \sum_{k=0}^{\ell-1}p_k(a_0) &&\text{by the induction hypothesis,}\\
 &= p_\ell(a_0) &&\text{by~\eqref{eqn pn}.}
\end{align*}

\item From~\eqref{eqn basics 2},
we have
$$\sum_{k=0}^{n-1}a_k\linebreak+\rho = a_0 a_{n-1} .$$
Therefore,
\begin{align*}
\rho &= a_0a_{n-1} - \sum_{k=0}^{n-1}a_k
\\
&= a_0 p_{n-1}(a_0) - \sum_{k=0}^{n-1}p_k(a_0) 
&& \text{by Theorem~\ref{theorem strategy with n cuts}-\ref{theorem strategy with n cuts item ai},}
\\
&= p_n(a_0) && \text{by~\eqref{eqn pn}.}
\end{align*}

We show $a_0 \geq a_n$ by contradiction.
Suppose $a_0 < \alpha_n$.
Then,
by Lemma~\ref{lemma properties pn item sign of pn},
there exists an $i\in\mathbb{N}$ such that $0 \leq i < n$ and $p_i(a_0) < 0$.
Hence,
$a_i = p_i(a_0) < 0$
by Theorem~\ref{theorem strategy with n cuts}-\ref{theorem strategy with n cuts item ai}.
This is impossible since all the $a_i$'s are such that $1 \leq a_i \leq \rho$.
Therefore,
$a_0 \geq \alpha_n$.
Moreover,
$a_0 \neq \alpha_n$
since $p_n(a_0) = \rho \geq 1$,
whereas $p_n(\alpha_n) = 0$ by the definition of $\alpha_n$.
Finally,
this solution is unique since $\alpha_n$ is the largest real root of $p_n$,
and the leading coefficient of $p_n$ is positive.

\item As we explained in the introduction,
$a_n = \rho$.
Also,
by Theorem~\ref{theorem strategy with n cuts}-\ref{theorem strategy with n cuts item a0},
$p_n(a_0) = \rho$.

\item This follows directly from the discussion preceding
Theorem~\ref{theorem strategy with n cuts}.
\qed
\end{enumerate}

From Theorem~\ref{theorem strategy with n cuts},
the optimal strategy $f_n$ is uniquely defined for each $n$.
However, this still leaves an infinite number of possibilities for the 
optimal strategy (one for each $n$). We aim to find, for a given $\rho$,
what value of $n$ leads to the optimal strategy.
Theorem~\ref{theorem optimal n}
gives a criterion for the optimal $n$ in terms of $\rho$
together with a formula that enables to compute this optimal $n$ in $O(1)$ time.
\begin{theorem}\
\label{theorem optimal n}
\begin{enumerate}
\item\label{theorem strategy with n cuts item best strategy}
For a given $\rho$,
if $n\in\mathbb{N}$ is such that 
\begin{align}
\label{condition a0 well defined}
p_n(\alpha_{n+1}) \leq \rho < p_n(\alpha_{n+2})  ,
\end{align}
then $f_n$ is the optimal strategy
and $\alpha_{n+1} \leq a_0 < \alpha_{n+2}$.

\item
For all $n\in\mathbb{N}$,
\begin{align}
\label{bounds pn alpha n+1 n+2}
2^n \leq p_n(\alpha_{n+1}) \leq \rho < p_n(\alpha_{n+2}) \leq 2^{n+2}  .
\end{align}
\end{enumerate}
\end{theorem}
Notice that the criterion in Theorem~\ref{theorem optimal n}-\ref{theorem strategy with n cuts item best strategy}
covers all possible values of $\rho$ since $p_0(\alpha_1) = 1$ by~\eqref{formula alpha n}
and $p_n(\alpha_{n+2}) = p_{n+1}(\alpha_{n+2})$
by Proposition~\ref{proposition properties pn}-\ref{proposition properties pn item solutions pn+1 = pn}
(see below).

To prove Theorem~\ref{theorem optimal n},
we need the following proposition.
\begin{proposition}
\label{proposition properties pn}
The following properties are true for all $n \in \mathbb{N}$.
\begin{enumerate}
\item\label{proposition properties pn item solutions pn+1 = pn}
$p_{n+1}(x) = p_n(x)$
if and only if
$p_{n+2}(x)=0$.

\item\label{proposition properties pn item alpha n increasing}
$0\leq \alpha_n < \alpha_{n+1} < 4$.

\item\label{proposition properties pn item n < n+1}
If $x \geq \alpha_{n+2}$,
then $p_{n+1}(x) \geq p_n(x)$.

\item\label{proposition properties pn item sign of pn+1 - pn}
For all $x\in\mathbb{R}$,
if $\alpha_{n+1} < x < \alpha_{n+2}$,
then $p_{n+1}(x) < p_n(x)$.

\item
\begin{align}
\label{formula pn alpha n+1}
p_n(\alpha_{n+1}) &= \alpha_{n+1}^{(n+1)/2} \\
\label{formula pn alpha n+2}
p_n(\alpha_{n+2}) &= \alpha_{n+2}^{(n+2)/2}
\end{align}
\end{enumerate}
\end{proposition}

\proof\
\begin{enumerate}
\item
\begin{align}
p_{n+1}(x) &= p_n(x) \\
x(p_{n+1}(x)-p_n(x)) &= 0 \\
p_{n+2}(x) &= 0
\end{align}

\item This follows directly from~\eqref{formula alpha n}.

\item Since $\alpha_{n+2}$ is the largest real root of $p_{n+2}$,
$p_{n+2}$ is strictly increasing on $[\alpha_{n+2},\infty)$.
Moreover,
since $x \geq \alpha_{n+2}$,
we have
\begin{align}
p_{n+2}(x) & \geq 0 \\
x(p_{n+1}(x) - p_n(x)) & \geq 0 \\
p_{n+1}(x) &\geq p_n(x)
\end{align}

\item
By \eqref{formula alpha n},
if $\alpha_{n+1} < x < \alpha_{n+2}$,
then
$$4\cos^2\!\left(\frac{\pi}{n+3}\right) < x < 4\cos^2\!\left(\frac{\pi}{n+4}\right) .$$
Hence,
since $\frac{\pi}{n+3} < \frac{2\pi}{n+4}$,
we have
$$4\cos^2\!\left(\frac{2\pi}{(n+2)+2}\right) < 4\cos^2\!\left(\frac{\pi}{n+3}\right) < x < 4\cos^2\!\left(\frac{\pi}{(n+2)+2}\right) .$$
Together with \eqref{factorization pn},
this implies that $p_{n+2}(x) < 0$.
Thus,
\begin{align*}
p_{n+2}(x) &< 0\\
x(p_{n+1}(x)-p_n(x)) &< 0\\
p_{n+1}(x) &< p_n(x) .
\end{align*}

\item
Notice that \eqref{formula pn alpha n+1}
is a direct consequence of \eqref{formula pn alpha n+2}
since $p_n(\alpha_{n+2}) = p_{n+1}(\alpha_{n+2})$
by Proposition~\ref{proposition properties pn}-\ref{proposition properties pn item solutions pn+1 = pn}.

We now show \eqref{formula pn alpha n+2}
by proving
\begin{align}
\label{eqn for proof}
p_n(\alpha_{n+2}) = \alpha_{n+2}^{i+1}\,\frac{p_{n-i}(\alpha_{n+2})}{p_i(\alpha_{n+2})} \qquad (0\leq i \leq n)
\end{align}
by induction on $i$.
If $i=0$,
then
$$p_n(\alpha_{n+2}) = \alpha_{n+2}^{0+1}\,\frac{p_{n-0}(\alpha_{n+2})}{p_0(\alpha_{n+2})} .$$
If $i = 1$,
then
\begin{align*}
p_n(\alpha_{n+2}) &= p_{n+1}(\alpha_{n+2}) &&\text{by Proposition~\ref{proposition properties pn}-\ref{proposition properties pn item solutions pn+1 = pn},} \\
&= \alpha_{n+2}(p_n(\alpha_{n+2})-p_{n-1}(\alpha_{n+2})) .
\end{align*}
Therefore,
\begin{align*}
p_n(\alpha_{n+2}) &= \alpha_{n+2}\frac{p_{n-1}(\alpha_{n+2})}{\alpha_{n+2}-1} \\
&= \alpha_{n+2}\frac{p_{n-1}(\alpha_{n+2})}{\frac{1}{\alpha_{n+2}}p_1(\alpha_{n+2})} \\
&= \alpha_{n+2}^{1+1}\frac{p_{n-1}(\alpha_{n+2})}{p_1(\alpha_{n+2})} .
\end{align*}

Suppose that~\eqref{eqn for proof} is true for $i=\ell-1$ and $i = \ell < n$,
we now prove it for $i = \ell+1$.
\begin{align*}
p_{n-(\ell-1)}(\alpha_{n+2}) &= \alpha_{n+2}(p_{n-\ell}(\alpha_{n+2})-p_{n-(\ell+1)}(\alpha_{n+2})) \\
\frac{p_{\ell-1}(\alpha_{n+2})}{\alpha_{n+2}^{(\ell-1)+1}}p_n(\alpha_{n+2}) &= \alpha_{n+2}\left(\frac{p_\ell(\alpha_{n+2})}{\alpha_{n+2}^{\ell+1}}p_n(\alpha_{n+2})-p_{n-(\ell+1)}(\alpha_{n+2})\right) \\
(p_\ell(\alpha_{n+2})-p_{\ell-1}(\alpha_{n+2}))p_n(\alpha_{n+2}) &= \alpha_{n+2}^{\ell+1}p_{n-(\ell+1)}(\alpha_{n+2}) \\
\frac{p_{\ell+1}(\alpha_{n+2})}{\alpha_{n+2}}p_n(\alpha_{n+2}) &= \alpha_{n+2}^{\ell+1}p_{n-(\ell+1)}(\alpha_{n+2}) \\
p_n(\alpha_{n+2}) &= \alpha_{n+2}^{(\ell+1)+1}\,\frac{p_{n-(\ell+1)}(\alpha_{n+2})}{p_{\ell+1}(\alpha_{n+2})} ,
\end{align*}
where the second equality comes from the induction hypothesis.
That completes the proof of~\ref{eqn for proof}.

Now,
by taking $i=n$ in~\ref{eqn for proof},
we find
\begin{align*}
p_n(\alpha_{n+2}) &= \alpha_{n+2}^{n+1}\,\frac{p_{n-n}(\alpha_{n+2})}{p_n(\alpha_{n+2})} ,\\
p_n^2\!(\alpha_{n+2}) &= \alpha_{n+2}^{n+1}\,p_0(\alpha_{n+2}) ,\\
p_n^2\!(\alpha_{n+2}) &= \alpha_{n+2}^{n+2} ,\\
p_n(\alpha_{n+2}) &= \alpha_{n+2}^{\frac{n+2}{2}} .
\end{align*}
\qed
\end{enumerate}

We now prove Theorem~\ref{theorem optimal n}.
\proof(Theorem~\ref{theorem optimal n})\
\begin{enumerate}
\item Consider the strategy $f_n$.
Since $\alpha_n$ is the largest real root of $p_n$,
$p_n$ is strictly increasing on $[\alpha_n,\infty)$.
Moreover,
by Proposition~\ref{proposition properties pn}-\ref{proposition properties pn item alpha n increasing},
$\alpha_n < \alpha_{n+1}$.
Therefore,
by Theorem~\ref{theorem strategy with n cuts}-\ref{theorem strategy with n cuts item a0}
and since $p_n(\alpha_{n+1}) \leq \rho < p_n(\alpha_{n+2})$ by the hypothesis,
we have
\begin{align}
\label{ineq proof contr 0}
\alpha_{n+1} \leq a_0 < \alpha_{n+2}
\end{align}

We first prove that $CR(f_n) \leq CR(f_m)$ for all $m < n$ by contradiction.
Suppose that there exists an $m < n$
such that $CR(f_m) < CR(f_n)$.
By Theorem~\ref{theorem strategy with n cuts}-\ref{theorem strategy with n cuts item a0},
there exists an $a_0'$ such that $a_0' > \alpha_m$ and $p_m(a_0') = \rho$.
Moreover,
since $CR(f_m) < CR(f_n)$ by the hypothesis,
then
$2a_0' + 1 < 2a_0+1$
by Theorem~\ref{theorem strategy with n cuts}-\ref{theorem strategy with n cuts item cost}.
Therefore,
\begin{align}
\label{ineq proof contr 1}
\alpha_m < a_0' < a_0  .
\end{align}
Also,
since $m < n$,
then $m + 2 \leq n+1$.
Thus,
by~\eqref{ineq proof contr 0}
and since the $\alpha_n$'s are increasing with respect to $n$
(see Proposition~\ref{proposition properties pn}-\ref{proposition properties pn item alpha n increasing}),
$\alpha_i \leq \alpha_{n+1} \leq a_0$
for all $m+2 \leq i \leq n+1$.
Hence,
by repeated applications of Proposition~\ref{proposition properties pn}-\ref{proposition properties pn item n < n+1},
we find
\begin{align}
\label{ineq proof contr 2}
p_m(a_0) \leq p_{m+1}(a_0) \leq p_{m+2}(a_0) \leq \ldots \leq p_{n-1}(a_0) \leq p_n(a_0)  .
\end{align}
But then,
\begin{align*}
\rho &= p_m(a_0') \\
&< p_m(a_0) && \text{by~\eqref{ineq proof contr 1} 
and since $p_m$ is increasing on $[\alpha_m,\infty)$,}\\
&\leq p_n(a_0) && \text{by~\eqref{ineq proof contr 2},}\\
&= \rho  ,
\end{align*}
which is a contradiction.
Consequently,
$CR(f_n) < CR(f_m)$ for all $m < n$.

We now prove that
$CR(f_n) \leq CR(f_{n'})$ for all $n' > n$ by contradiction.
Suppose that there exists an $n' > n$
such that $CR(f_{n'}) < CR(f_n)$.
By Theorem~\ref{theorem strategy with n cuts}-\ref{theorem strategy with n cuts item a0},
there exists an $a_0'$ such that $a_0' > \alpha_{n'}$ and $p_{n'}(a_0') = \rho$.
Moreover,
since $CR(f_{n'}) < CR(f_n)$ by the hypothesis,
then
$2a_0' + 1 < 2a_0+1$
by Theorem~\ref{theorem strategy with n cuts}-\ref{theorem strategy with n cuts item cost}.
Therefore,
by~\eqref{ineq proof contr 0}
and since the $\alpha_n$'s are strictly increasing with respect to $n$
(see Proposition~\ref{proposition properties pn}-\ref{proposition properties pn item alpha n increasing}),
\begin{align}
\label{ineq proof contr 3}
\alpha_n < \alpha_{n'} < a_0' < a_0 < \alpha_{n+2}.
\end{align}
Moreover,
since $n'>n$,
\eqref{ineq proof contr 3} implies $n' = n+1$.
But then,
\begin{align*}
\rho &= p_{n'}(a_0') \\
&= p_{n+1}(a_0') \\
&< p_n(a_0') && \text{by~\eqref{ineq proof contr 3} and Proposition~\ref{proposition properties pn}-\ref{proposition properties pn item sign of pn+1 - pn},}\\
&< p_n(a_0) && \text{by~\eqref{ineq proof contr 3} and since $p_n$ is increasing on $[\alpha_n,\infty)$,}\\
&= \rho  ,
\end{align*}
which is a contradiction.
Consequently,
$CR(f_n) < CR(f_{n'})$ for all $n' > n$.

\item
We first prove that
\begin{align}
\label{eqn thm proof 1}
2\cos^{n+1}\!\left(\frac{\pi}{n+3}\right) \geq 1
\end{align}
for all $n \geq 0$.
We can easily verify \eqref{eqn thm proof 1} for $n = 0$ and $n = 1$.
We provide a general proof for $n \geq 2$.
By elementary calculus,
we have $\cos(x) \geq 1-x^2/2$
for all $0 \leq x \leq \pi/4$.
We have $0\leq \pi/(n+3) \leq \pi/4$ because $n\geq 2$.
Hence,
if we can prove
\begin{align}
\label{eqn thm proof 2}
g(n) = 2\left(1-\frac{1}{2}\left(\frac{\pi}{n+3}\right)^2\right)^{n+1} \geq 1
\end{align}
for $n \geq 2$,
then we are done.
We have $g'(n) = g(n)h(n)$,
where
$$h(n) = \log\left(1-\frac{1}{2}\left(\frac{\pi}{n+3}\right)^2\right)+\frac{(n+1)\pi^2}{(n+3)^3\left(1-\frac{1}{2}\left(\frac{\pi}{n+3}\right)^2\right)} .$$
The function $g$ is positive for $n \geq 2$.
We can prove by elementary calculus that $h$ also is positive for $n \geq 2$.
Therefore,
we conclude that $g'$ is strictly positive,
and hence,
that $g$ is strictly increasing for $n \geq 2$.
Thus,
since $g(2) > 1$,
then \eqref{eqn thm proof 2} is true for $n \geq 2$ and
the proof of~\eqref{eqn thm proof 1} is complete.

We now prove \eqref{bounds pn alpha n+1 n+2}.
\begin{align*}
2^n &\leq 2^{n+1}\cos^{n+1}\!\left(\pi/(n+3)\right) && \text{by~\eqref{eqn thm proof 1},}\\
&= \alpha_{n+1}^{(n+1)/2} && \text{by~\eqref{factorization pn},}\\
&= p_n(\alpha_{n+1}) && \text{by~\eqref{formula pn alpha n+1},}\\
&\leq \rho && \text{by~\eqref{condition a0 well defined},}\\
&< p_n(\alpha_{n+2}) && \text{by~\eqref{condition a0 well defined},}\\
&= \alpha_{n+2}^{(n+2)/2} && \text{by~\eqref{formula pn alpha n+2},}\\
&= 2^{n+2}\cos^{n+2}\!\left(\pi/(n+4)\right) 
&& \text{by~\eqref{factorization pn},}\\
&\leq 2^{n+2} && \text{since $0 < \cos(\pi/(n+4)) < 1$.}
\end{align*}
\qed
\end{enumerate}

From~\eqref{condition a0 well defined},
there is only one possible optimal value for $n$.
From Theorem~\ref{theorem strategy with n cuts},
once we are given a $\rho$ and an $n$,
there is only one possible optimal strategy.
Therefore,
we have the following corollary.
\begin{corollary}
For a given $\rho$,
there exists a unique optimal strategy that is monotonic and periodic.
\end{corollary}

By~\eqref{bounds pn alpha n+1 n+2},
it suffices to examine two values to find the optimal $n$,
namely $\lfloor \log_2\rho\rfloor -1$ and $\lfloor \log_2\rho\rfloor$.
To compute the optimal $n$, let $n = \lfloor \log_2\rho\rfloor$
and let $\gamma = 2\cos\left(\frac{\pi}{n+3}\right)$.
If $n+1 \leq \log_{\gamma}\rho$,
then $n$ is optimal.
Otherwise,
take $n = \lfloor \log_2\rho\rfloor - 1$.
By Theorem~\ref{theorem optimal n},
this gives us the optimal $n$ in $O(1)$ time.

Now that we know the optimal $n$,
we need to compute $a_i$ for each $0 \leq i < n$.
Suppose that we know $a_0$.
By~\eqref{eqn def pi},
and Theorems~\ref{theorem strategy with n cuts}-\ref{theorem strategy with n cuts item ai} and~\ref{theorem strategy with n cuts}-\ref{theorem strategy with n cuts item an},
$a_1 = p_1(a_0) = a_0(a_0-1)$
and $a_i = a_0(p_{i-1}(a_0)-p_{i-2}(a_0)) = a_0(a_{i-1}-a_{i-2})$
for $2 \leq i \leq n$.
Therefore,
given $a_0$,
each $a_i$ can be computed in $O(1)$ time for $1\leq i \leq n$.
It remains to show how to compute $a_0$ efficiently.
Since $f_n$ is defined by $n$ values,
$\Omega(n) = \Omega(\log\rho)$ time is necessary to compute $f_n$.
Hence, if we can compute $a_0$ in $O(1)$ time,
then our algorithm is optimal.

By Theorem~\ref{theorem strategy with n cuts}-\ref{theorem strategy with n cuts item a0},
for a given $n$,
we need to solve a polynomial equation of degree $n+1$ to find the value of $a_0$.
By Galois theory, 
this cannot be done by radicals if $n+1 > 4$.
Moreover,
the degree of the $p_n$'s is unbounded,
so $a_0$ cannot be computed exactly in general.
Theorem~\ref{theorem approximate solution}
explains how and why numerical methods can be used efficiently to address this issue.
\begin{theorem}
\label{theorem approximate solution}
Take $\rho$ and $n$ such that $f_n$ is optimal for $\rho$.
\begin{enumerate}
\item\label{theorem approximate solution item alpha n+2}
Let $a_0^* \in \mathbb{R}$ be such that $\alpha_{n+1} \leq a_0 < a_0^* \leq \alpha_{n+2}$
and define $f_n^*$ by
$$f_n^*(i) = \begin{cases}
p_i(a_0^*)\Dmin & \text{if $0\leq i < n$,} \\ 
\rho\Dmin & \text{if $i \geq n$.}
\end{cases}$$
Then $\left|CR(f_n)-CR(f_n^*)\right| \leq 7^3\,(n+4)^{-3}$.

\item\label{theorem approximate solution item good interval}
The polynomial $p_n$ is strictly increasing on $[\alpha_{n+1},\alpha_{n+2})$
and $|\alpha_{n+2}-\alpha_{n+1}| \leq 7^3\,(n+4)^{-3}/2$.
\end{enumerate}
\end{theorem}

\proof\
\begin{enumerate}
\item Let $a_i^* = p_i(a_0^*)$ for $0\leq i < n$.
We first prove that $CR(f_n^*) = 2a_0^*+1$.
By Theorems~\ref{theorem strategy with n cuts}
and~\ref{theorem optimal n}-\ref{theorem strategy with n cuts item best strategy},
there is a $\rho^*\in\mathbb{R}$ such that
$p_n(\alpha_{n+1}) \leq \rho < \rho^* \leq p_n(\alpha_{n+2})$,
$p_n(a_0^*) = \rho^*$
and $f_n^*$ is optimal for $\rho^*$.
By Theorem~\ref{theorem strategy with n cuts} and the discussion preceding it,
we have
\begin{align*}
\sup_{D \in [\Dmin,a_0^*\Dmin]} \frac{1}{D}\phi(f_n^*,D) &= 2a_0^*+1  ,\\
\sup_{D \in (a_i^*\Dmin,a_{i+1}^*\Dmin]} \frac{1}{D}\phi(f_n^*,D) &= 1 + 2\sum_{k=0}^{i+1}\frac{a_k^*}{a_i^*} 
&& (0 \leq i \leq n-2) \\
&= 2a_0^*+1 
&& (0 \leq i \leq n-2)  ,\\
\sup_{D \in (a_{n-1}^*\Dmin,\rho\Dmin]} \frac{1}{D}\phi(f_n^*,D) &= 1 + 2\sum_{k=0}^{n-1}\frac{a_k^*}{a_{n-1}^*}+2\frac{\rho}{a_{n-1}^*} \\
&< 1 + 2\sum_{k=0}^{n-1}\frac{a_k^*}{a_{n-1}^*}+2\frac{\rho^*}{a_{n-1}^*} \\
&= 2a_0^*+1  .
\end{align*}
This establishes that $CR(f_n^*) = 2a_0^*+1$.
Therefore,
\begin{align*}
& \phantom{=} \left|CR(f_n)-CR(f_n^*)\right| \\
&= \left|(2a_0+1)-(2a_0^*+1)\right| 
&& \text{by Theorem~\ref{theorem strategy with n cuts}-\ref{theorem strategy with n cuts item cost} and since $CR(f_n^*) = 2a_0^*+1$,}\\
&= 2(a_0^*-a_0) \\
&\leq 2(\alpha_{n+2}-\alpha_{n+1}) 
&& \text{by the hypothesis and Theorem~\ref{theorem optimal n}-\ref{theorem strategy with n cuts item best strategy},} \\
&= 8\left(\cos^2\!\left(\frac{\pi}{n+4}\right) - \cos^2\!\left(\frac{\pi}{n+3}\right)\right)
&& \text{by~\eqref{factorization pn}.}\\
&\leq 7^3\, (n+4)^{-3} && \text{by elementary calculus.}
\end{align*}

\item
Since $\alpha_n$ is the largest real root of $p_n$,
$p_n$ is strictly increasing on $[\alpha_n,\infty)$.
Since the $\alpha_n$'s are strictly increasing with respect to $n$
(see Proposition~\ref{proposition properties pn}-\ref{proposition properties pn item alpha n increasing}),
$p_n$ is strictly increasing on $[\alpha_{n+1},\alpha_{n+2}) \subset [\alpha_n,\infty)$.

The inequality $|\alpha_{n+2}-\alpha_{n+1}| \leq 7^3\,(n+4)^{-3}/2$
follows directly from the proof of Theorem~\ref{theorem approximate solution}-\ref{theorem approximate solution item alpha n+2}
\qed
\end{enumerate}

We now explain how to compute $a_0$.
We know what is the optimal $n$ for a given $\rho$.
From~\eqref{factorization pn} and Theorem~\ref{theorem optimal n}-\ref{theorem strategy with n cuts item best strategy}, 
$n$ satisfies $n+1 \leq 4$
if and only if $\rho \leq 32\cos^5(\pi/7) \approx 18.99761$.
In this case,
$p_n(x) = \rho$ is a polynomial equation of degree at most $4$.
Hence,
by Theorem~\ref{theorem strategy with n cuts}-\ref{theorem strategy with n cuts item a0}
and elementary algebra,
$a_0$ can be computed exactly and in $O(1)$ time.
Otherwise,
let $\varepsilon > 0$ be a given tolerance.
We explain how to find a solution $f_{opt}^*$,
such that $CR(f_{opt}^*) \leq CR(f_{opt}) + \varepsilon$.

If $n \geq 7\varepsilon^{-1/3}-4$,
then by Theorem~\ref{theorem approximate solution},
it suffices to take $a_0 = \alpha_{n+2}$
to compute an $\varepsilon$-approximation of the optimal strategy.
But $\alpha_{n+2} = 4\cos^2\!(\pi/(n+4))$ by~\eqref{factorization pn}.
Hence,
$a_0$ can be computed in $O(1)$ time and thus,
an $\varepsilon$-approximation of the optimal strategy
can be computed in $\Theta(n) = \Theta(\log\rho)$ time.
Otherwise,
if $4 \leq n < 7\varepsilon^{-1/3}-4$,
then we have to use numerical methods to find the value of $a_0$.
By Theorem~\ref{theorem optimal n}-\ref{theorem strategy with n cuts item best strategy},
we need to solve $p_n(x) = \rho$ for $x\in[\alpha_{n+1},\alpha_{n+2})$.
However,
by Theorem~\ref{theorem approximate solution}-\ref{theorem approximate solution item good interval},
$|\alpha_{n+2} - \alpha_{n+1}| < 7^3(n+4)^{-3}/2$
and $p_n$ is strictly increasing on this interval.
Hence,
usual root-finding algorithms behave well
on this problem.

Hence,
if $n < 4$ or $n \geq 7\varepsilon^{-1/3}-4$,
then our algorithm is optimal.
When $4 \leq n < 7\varepsilon^{-1/3}-4$,
then our algorithm's computation time
is as fast as the fastest root-finding algorithm.

It remains to provide bounds on $CR(f_n)$ for an optimal $n$;
we present exact bounds in Theorem~\ref{theorem CR}.
\begin{theorem}\
\label{theorem CR}
\begin{enumerate}
\item\label{theorem CR item bounds CR}
The strategy $f_0$ is optimal for a given $\rho $ if and only if $1\leq \rho < 2$.
In this case,
$CR(f_0) = 2\rho+1$.

Otherwise,
if $f_n$ is optimal for a given $\rho$ ($n\geq 1$),
then
\begin{align}
\label{almost exact bounds}
8\cos^2\!\left(\frac{\pi}{\lceil\log_2 \rho\rceil + 2}\right) + 1 \leq CR(f_n) \leq 8\cos^2\!\left(\frac{\pi}{\lfloor\log_2 \rho\rfloor + 4}\right) + 1  .
\end{align}

\item\label{theorem CR item infinity}
For a fixed $\Dmin$,
when $\Dmax \rightarrow \infty$,
$f_{opt} \rightarrow f_{\infty}$,
where
$f_{\infty}(i) = (2i+4)2^i \Dmin$ ($i \geq 0$)
and $CR(f_{\infty}) = 9$.
\end{enumerate}
\end{theorem}

\proof\
\begin{enumerate}
\item The first statement is a direct consequence of
Theorem~\ref{theorem optimal n}-\ref{theorem strategy with n cuts item best strategy},
\eqref{formula alpha n}
and Theorem~\ref{theorem strategy with n cuts}-\ref{theorem strategy with n cuts item cost}.

Otherwise,
if $f_n$ is optimal ($n \geq 1$),
then $2^n \leq \rho < 2^{n+2}$ by~\eqref{bounds pn alpha n+1 n+2}.
Therefore,
\begin{align*}
\log_2(\rho) -2 &< n \leq \log_2(\rho) ,\\
\left\lceil\log_2(\rho)\right\rceil -2 &\leq n \leq \lfloor\log_2(\rho)\rfloor ,
\end{align*}
since $n$ is an integer.
Moreover,
$CR(f_n) = 2a_0 + 1$
by Theorem~\ref{theorem strategy with n cuts}-\ref{theorem strategy with n cuts item cost}.
Hence,
by Theorem~\ref{theorem optimal n}-\ref{theorem strategy with n cuts item best strategy},
\eqref{formula alpha n}
and the previous derivation,
we get
\begin{align*}
2\alpha_{n+1}+1 &\leq CR(f_n) \leq 2\alpha_{n+2}+1 ,\\
8\cos^2\!\left(\frac{\pi}{\lceil\log_2\rho\rceil+1}\right)+1 &\leq CR(f_{opt}) < 8\cos^2\!\left(\frac{\pi}{\lfloor\log_2\rho\rfloor+4}\right)+1 .
\end{align*}

\item Let $f_n$ be the optimal strategy for $\rho$
and suppose that $\Dmin$ is fixed.
When $\Dmax \rightarrow \infty$,
then $\rho \rightarrow \infty$
and then,
$n \rightarrow \infty$ by~\eqref{bounds pn alpha n+1 n+2}.
Hence,
by Theorem~\ref{theorem optimal n}-\ref{theorem strategy with n cuts item best strategy}
and~\eqref{factorization pn},
$$4 = \lim_{n\rightarrow \infty} \alpha_{n+1} \leq \lim_{n\rightarrow \infty} a_0 \leq \lim_{n\rightarrow \infty} \alpha_{n+2} = 4 .$$

Let us now prove that
$$p_i(4) = (2i+4)2^i$$
for all $i \geq 0$.
We proceed by induction on $i$.
For the base case,
notice that $p_0(4) = 4 = (2(0)+4)2^0$
and $p_1(4) = 4(4-1) = (2(1)+4)2^1$.
Suppose that
\begin{align*}
p_{i-1}(4) &= (2(i-1)+4)2^{i-1} ,\\
p_i(4) &= (2i+4)2^i ,
\end{align*}
(for an $i \geq 1$)
and let us prove that $p_{i+1}(4) = (2(i+1)+4)2^{i+1}$.
\begin{align*}
p_{i+1}(4) &= 4(p_i(4) - p_{i-1}(4)) \\
& = 4\left( (2i+4)2^i - (2(i-1)+4)2^{i-1} \right) &&\text{by the induction hypothesis,}\\
& = (2(i+1)+4)2^{i+1} \enspace.
\end{align*}

Thus,
when $\Dmax \rightarrow \infty$,
$a_i = p_i(a_0) = p_i(4) = (2i+4)2^i$
by Theorem~\ref{theorem strategy with n cuts}-\ref{theorem strategy with n cuts item ai}
and since $p_i$ is continuous.
Hence,
$f_n \rightarrow f_{\infty}$.
\qed
\end{enumerate}

The competitive cost of the optimal strategy is $2a_0+1$
by Theorem~\ref{theorem strategy with n cuts}-\ref{theorem strategy with n cuts item cost}.
Theorem~\ref{theorem CR}-\ref{theorem CR item bounds CR}
gives nearly tight bounds on $2a_0+1$.
Notice that when $\rho = 1$,
i.e., when $D$ is known,
then $2a_0+1 = 3$ which corresponds to the optimal strategy in this case.
From the Taylor series expansion of $\cos^2\!(\cdot)$ 
and Theorem~\ref{theorem CR}-\ref{theorem CR item bounds CR},
we have $CR(f_n) = 9-O(1/\log^2\rho)$
for an optimal $n$.
This is consistent with L\'opez-Ortiz and Schuierer' result
(see~\cite{DBLP:journals/tcs/Lopez-OrtizS01}),
although our result~\eqref{almost exact bounds} is exact.

Letting $\rho \rightarrow \infty$
corresponds to not knowing any upper bound on $D$.
Thus, Theorem~\ref{theorem CR}-\ref{theorem CR item infinity}
provides an alternate proof to the competitive ratio of 9 shown by 
Baeza-Yates et al.~\cite{DBLP:journals/iandc/Baeza-YatesCR93}.
From Theorems~\ref{theorem optimal n} and~\ref{theorem CR},
the optimal solution for a given $\rho$ is unique.
This optimal solution tends towards $f_{\infty}$, suggesting
that $f_{\infty}$ is the canonical optimal strategy
when no upper bound is given
(rather than the power of two strategy).

In this section,
we proved the following theorem.
\begin{theorem}
\label{theorem algorithm}
Let $\Dmin$, $\Dmax$ and $\varepsilon>0$ be given,
where $0 < \Dmin \leq D \leq \Dmax$.
\begin{itemize}
\item If $\rho= \Dmax/\Dmin  \leq 32\cos^5(\pi/7) \approx 18.99761$,
Algorithm~\ref{algorithm optimal strategy} computes,
in $O(1)$ time,
the exact optimal strategy $f_{opt}$.

\item Otherwise,
Algorithm~\ref{algorithm optimal strategy} computes
a strategy $f_{opt}^*$ such that $CR(f_{opt}^*) \leq CR(f_{opt}) + \varepsilon$.
\begin{itemize}
\item If $n \geq 7\varepsilon^{-1/3}-4$,
Algorithm~\ref{algorithm optimal strategy}
computes $f_{opt}^*$ in $O(n)=O(\log \rho)$ time.

\item Otherwise,
the time needed for Algorithm~\ref{algorithm optimal strategy} to compute $f_{opt}^*$
is equal to the time needed for the fastest root-finding algorithm to solve
$p_n(x) = \rho$ on $[\alpha_{n+1},\alpha_{n+2})$.
\end{itemize}
\end{itemize}
\end{theorem}

\section{Maximal Reach Problem}
\label{section maximal reach}

In this section,
we explain how to solve the maximal reach problem
using the result from Section~\ref{section searching bounded line}.
Given a competitive ratio $R$ and a lower bound $\Dmin$ on $D$,
the \emph{maximal reach problem}
is to identify the largest bound $\Dmax$ such that there exists a search strategy
that finds any target within distance $\Dmin \leq D \leq \Dmax$
with competitive ratio at most $R$.

We have $a_0 = \frac{1}{2}(R-1)$
by Theorem~\ref{theorem strategy with n cuts}-\ref{theorem strategy with n cuts item cost}.
Therefore,
by Theorem~\ref{theorem optimal n}-\ref{theorem strategy with n cuts item best strategy}
and~\eqref{formula alpha n},
\begin{align*}
\alpha_{n+1} &\leq a_0 < \alpha_{n+2} ,\\
4\cos^2\!\left(\frac{\pi}{n+3}\right) &\leq a_0 < 4\cos^2\!\left(\frac{\pi}{n+4}\right) ,\\
\cos\left(\frac{\pi}{n+3}\right) &\leq \frac{\sqrt{a_0}}{2} < \cos\left(\frac{\pi}{n+4}\right) ,\\
\frac{\pi}{\arccos\left(\frac{\sqrt{a_0}}{2}\right)} - 4 &< n \leq \frac{\pi}{\arccos\left(\frac{\sqrt{a_0}}{2}\right)} - 3 ,
\end{align*}
from which we find
$$n = \left\lfloor \frac{\pi}{\arccos\left(\frac{\sqrt{a_0}}{2}\right)}\right\rfloor - 3 = \left\lfloor \frac{\pi}{\arccos\left(\frac{\sqrt{R-1}}{2\sqrt{2}}\right)}\right\rfloor - 3$$
since $n$ is an integer.

Consequently,
$$\Dmax = p_n(a_0) \Dmin,$$
which can be computed in $O(n)$ by the definition of the $p_n$'s.

\section{Searching on $m$ Bounded Concurrent Rays}
\label{section searching m rays}

For $m \geq 2$,
if we know $D$,
then the optimal strategy has a competitive cost of $1+2(m-1)$.
Indeed,
in the worst case,
we have to walk $2D$ on the first $m-1$ rays and then $D$ on the $m$-th ray.
When no upper bound is known,
Baeza-Yates et al.~\cite{DBLP:journals/iandc/Baeza-YatesCR93}
proved that the optimal strategy has a competitive cost of
$$1+2\frac{m^m}{(m-1)^{m-1}} .$$
There exist infinitely many strategies that achieve this optimal cost.
\begin{proposition}
\label{proposition infinite family m > 2}
All the strategies in the following family are optimal:
$$f_{a,b}(i) = (ai+b)\left(\frac{m}{m-1}\right)^i\Dmin ,$$
where $a \geq 0$
and
\begin{align}
\label{condition b}
\sup\{1,ma\} \leq b\leq \frac{\left(\frac{m^m}{(m-1)^{m-1}} - m^2\right)a+\frac{m}{m-1}\,\frac{m^m}{(m-1)^{m-1}}}{\frac{m^m}{(m-1)^{m-1}} - m} .
\end{align}
\end{proposition}
Notice that for $m=2$,
when $a$ and $b$ are respectively equal to their smallest allowed value,
then $f_{a,b}$ corresponds to the power of two strategy of Baeza-Yates et al.
(refer to~\cite{DBLP:journals/iandc/Baeza-YatesCR93}).
Moreover,
when $a$ is equal to its largest allowed value,
i.e. $a = 2$,
then $b = 4$ and $f_{a,b} = f_{\infty}$
(refer to Theorem~\ref{theorem CR}-\ref{theorem CR item infinity}).

\proof
Without loss of generality,
let $\Dmin = 1$.
We first explain why we need $a\geq 0$ and~\eqref{condition b}.

For any $a\geq 0$ and any $b\geq 0$,
the function $f_{a,b}$ is strictly increasing on $[0,\infty)$.
However,
to have $f_{a,b}(i) \geq 1$ for all $i \geq 0$,
we also need $b \geq 1$.
Indeed,
$f_{a,b}(0) = b$
and $f_{a,b}(i) \geq 1$ for all $i\geq 1$ implies
$$a \geq \frac{1}{i\left(\frac{m}{m-1}\right)^i}-\frac{b}{i}$$
for all $i \geq 1$.
Therefore,
we must have $a\geq 0$,
which we already knew.

We must keep control on the competitive cost of $f_{a,b}$.
The following two inequalities must be satisfied for all $D$:
\begin{align}
\label{family ineq 0}
1+2(m-1) \leq \frac{\phi(f_{a,b},D)}{D} \leq 1 + 2\frac{m^m}{(m-1)^{m-1}} .
\end{align}
The first inequality ensures that $f_{a,b}$ does not do better than the 
optimal strategy for the case where we know $D$
(refer to the discussion at the beginning of this section).
The second inequality ensures that $f_{a,b}$ is optimal for all $D$.
We consider the case where $1 = \Dmin \leq D < f_{a,b}(0)$ separately.
In this case,
we have
\begin{align*}
\frac{\phi(f_{a,b},D)}{D} &= \frac{2\sum\limits_{i=0}^{m-2}f_{a,b}(i) + D}{D} \\
&= \frac{1}{mD}\left(2(m-1)\left((b-a)\frac{m^m}{(m-1)^{m-1}}-m(b-am)\right) + D \right) .
\end{align*}
Therefore,
from~\eqref{family ineq 0},
we must have
$$1+2(m-1)\leq\lim_{D\rightarrow f_{a,b}(0)}\frac{\phi(f_{a,b},D)}{D} \leq \frac{\phi(f_{a,b},D)}{D} \leq \lim_{D\rightarrow 1}\frac{\phi(f_{a,b},D)}{D} \leq 1+2\frac{m^m}{(m-1)^{m-1}},$$
from which we get
\begin{align}
\label{family ineq 1}
\frac{2(m-1)}{mb}\left((b-a)\frac{m^m}{(m-1)^{m-1}}-m(b-am)\right) + 1 &\geq 1+2(m-1),\\
\label{family ineq 2}
\frac{2(m-1)}{m}\left((b-a)\frac{m^m}{(m-1)^{m-1}}-m(b-am)\right) + 1 &\leq 1+2\frac{m^m}{(m-1)^{m-1}}.
\end{align}
Equation~\eqref{family ineq 1}
leads to
$$b\geq \frac{\frac{m^m}{(m-1)^{m-1}} - m^2}{\frac{m^m}{(m-1)^{m-1}} - 2m} \, a .$$
Since
$$\frac{\frac{m^m}{(m-1)^{m-1}} - m^2}{\frac{m^m}{(m-1)^{m-1}} - 2m} \leq 0$$
and $b \geq 1$,
that condition is already satisfied.
Equation~\eqref{family ineq 2}
leads to
\begin{align*}
b\leq \frac{\left(\frac{m^m}{(m-1)^{m-1}} - m^2\right)a+\frac{m}{m-1}\,\frac{m^m}{(m-1)^{m-1}}}{\frac{m^m}{(m-1)^{m-1}} - m} .
\end{align*}

We now consider the general case where $f_{a,b}(j) \leq D < f_{a,b}(j+1)$
for a $j \geq 1$.
After simplification,
we get
\begin{align*}
\frac{\phi(f_{a,b},D)}{D} &= \frac{2\sum\limits_{i=0}^{(j+1)+(m-2)}f_{a,b}(i)+D}{D}\\
&= \frac{2(m-1)\left((aj+b)\left(\frac{m}{m-1}\right)^{m+j}-(b-am)\right)+D}{D} .
\end{align*}
Therefore,
from~\eqref{family ineq 0},
we must have
$$1+2(m-1) \leq \lim_{D\rightarrow f_{a,b}(j+1)}\frac{\phi(f_{a,b},D)}{D} \leq \frac{\phi(f_{a,b},D)}{D} \leq \lim_{D\rightarrow f_{a,b}(j)}\frac{\phi(f_{a,b},D)}{D} \leq 1+2\frac{m^m}{(m-1)^{m-1}},$$
from which we get
\begin{align}
\label{family ineq 3}
2\frac{(\frac{m}{m-1})^j\left((aj+b)\frac{m^m}{(m-1)^{m-1}} - m(a(j+1)+b)\right)-(m-1)(b-am)}{(a(j+1)+b)\left(\frac{m}{m-1}\right)^{j+1}}
&\geq  1+2(m-1),\\
\label{family ineq 4}
1+2\frac{m^m}{(m-1)^{m-1}} - \frac{2(m-1)}{aj+b}\left(\frac{m}{m-1}\right)^{-j}(b-am) &\leq 1+2\frac{m^m}{(m-1)^{m-1}}.
\end{align}

Equation~\eqref{family ineq 3}
leads to
$$b \geq-\frac{\left(\frac{m}{m-1}\right)^j\left(j\frac{m^m}{(m-1)^{m-1}}-(j+1)m\right) + m(m-1)}{\left(\frac{m}{m-1}\right)^j\left(\frac{m^m}{(m-1)^{m-1}}-m\right)-(m-1)} \, a.$$
Since
$$-\frac{\left(\frac{m}{m-1}\right)^j\left(j\frac{m^m}{(m-1)^{m-1}}-(j+1)m\right) + m(m-1)}{\left(\frac{m}{m-1}\right)^j\left(\frac{m^m}{(m-1)^{m-1}}-m\right)-(m-1)} \leq 0$$
and $b \geq 1$,
that condition is already satisfied.
Equation~\eqref{family ineq 4}
leads to $b \geq ma$.

Finally,
\begin{align*}
CR(f_{a,b}) &= \sup_{D \geq \Dmin} \frac{\phi(f_{a,b},D)}{D} \\
&= \lim_{j\rightarrow\infty}\lim_{D\rightarrow f_{a,b}(j)} \frac{2(m-1)\left((aj+b)\left(\frac{m}{m-1}\right)^{m+j}-(b-am)\right)+D}{D} \\
&= 9. & \qed
\end{align*}

When we are given an upper bound $\Dmax \geq D$,
the solution presented in Section~\ref{section searching bounded line}
partially applies to the problem of searching on $m$ concurrent bounded rays.
In this setting,
we start at the crossroads and we know that the target is on one of the $m$ rays
at a distance $D$ such that $\Dmin \leq D \leq \Dmax$.
Given a strategy $f(i)$,
we walk a distance of $f(i)$ on the $(i \bmod{m})$-th ray and go back to the crossroads.
We repeat for all $i \geq 0$ until we find the target.
As in the case where $m=2$,
we can suppose that is the solution is periodic and monotone
(refer to Section~\ref{section searching bounded line} 
or see Lemmas 2.1 and 2.2 in~\cite{DBLP:journals/tcs/Lopez-OrtizS01}).

Unfortunately,
we have not managed to push the analysis as far as in the case where $m=2$
because the expressions in the general case do not simplify as easily.
We get the following system of equations by
applying similar techniques as in Section~\ref{section searching bounded line}
\begin{align*}
1+2\sum_{k=0}^{i+m-1} \frac{a_k}{a_i} &= 1+2\sum_{k=0}^{m-2}a_k & 
(0 \leq i \leq n-m), \\
1+2\sum_{k=0}^{n-1}\frac{a_k}{a_i}+\frac{(i-(n-m))\rho}{a_i} &= 1+2\sum_{k=0}^{m-2}a_k & (n-m + 1 \leq i \leq n-1), 
\end{align*}
for $f_n$,
where
$$f_n(i) = \begin{cases} a_i\Dmin & \text{if $0\leq i < n$,} \\ 
\rho\Dmin & \text{if $i \geq n$.} \end{cases}$$
We prove in Theorem~\ref{theorem fundamental m rays}
that the solution to this system of equations can be obtained using
the following family of polynomials in $m-1$ variables,
where $\overline{x} = (x_0,x_1,...,x_{m-2})$
and $|\overline{x}| = x_0 + x_1 +...+ x_{m-2}$.
\begin{align*}
p_n(\overline{x}) &= x_n && (0 \leq n \leq m - 2)\\
p_{m-1}(\overline{x}) &= |\overline{x}|(x_0-1)\\
p_n(\overline{x}) &= |\overline{x}|(p_{n-(m-1)}(\overline{x}) - p_{n-m}(\overline{x})) && (n \geq m)
\end{align*}
In the rest of this section, for all $n\in\mathbb{N}$,
we let $\overline{\alpha}_n=(\alpha_{n,0},\alpha_{n,1},...,\alpha_{n,m-2})$ be 
the (real) solution to the system
\[
p_n(\overline{x}) = 0 , \quad
p_{n+1}(\overline{x}) = 0 , \quad
\ldots , \quad
p_{n+m-2}(\overline{x}) = 0
\]
such that
\begin{equation}
\label{constraint alpha n i}
0 \leq \alpha_{n,0} \leq \alpha_{n,1} \leq \cdots \leq \alpha_{n,m-2}
\end{equation}
and $|\overline{\alpha}_n|$ is maximized
(refer to Table~\ref{table examples alpha n}
for examples with $2 \leq m \leq 5$ and $0 \leq n \leq 6$).
\begin{table}
\begin{tabular}{r@{ $=$ }l|r@{ $=$ }l}
\multicolumn{2}{c|}{$m=2$} & \multicolumn{2}{c}{$m=3$}\\\hline
$\alpha_0$ & $0$ 
& $\overline{\alpha}_0$ & $(0,0)$ \\
$\alpha_1$ & $1$ 
& $\overline{\alpha}_1$ & $(0,0)$ \\
$\alpha_2$ & $2$ 
& $\overline{\alpha}_2$ & $(1,1)$ \\
$\alpha_3$ & $\frac{3+\sqrt{5}}{2}$ 
& $\overline{\alpha}_3$ & $\left(\frac{3}{2},\frac{3}{2}\right)$ \\
$\alpha_4$ & $3$ 
& $\overline{\alpha}_4$ & $\left(\frac{3+\sqrt{3}}{3},\frac{3+2\sqrt{3}}{3}\right)$ \\
$\alpha_5$ & $\frac{1}{3}\left(5+\frac{7^{2/3}}{\left(\frac{1}{2}\left(1+3i\sqrt{3}\right)\right)^{1/3}}+\left(\frac{7}{2}\left(1+3i\sqrt{3}\right)\right)^{1/3}\right)$ 
& $\overline{\alpha}_5$ & $\left(\frac{7+\sqrt{13}}{6},\frac{4+\sqrt{13}}{3}\right)$ \\
$\alpha_6$ & $2+\sqrt{2}$ 
& $\overline{\alpha}_6$ & $\left(\frac{15+3\sqrt{3}}{11},\frac{18+8\sqrt{3}}{11}\right)$ \\\hline
\multicolumn{2}{c|}{$m=4$} & \multicolumn{2}{c}{$m=5$} \\\hline
$\overline{\alpha}_0$ & $(0,0,0)$
& $\overline{\alpha}_0$ & $(0,0,0,0)$\\
$\overline{\alpha}_1$ & $(0,0,0)$
& $\overline{\alpha}_1$ & $(0,0,0,0)$\\
$\overline{\alpha}_2$ & $(0,0,0)$
& $\overline{\alpha}_2$ & $(0,0,0,0)$\\
$\overline{\alpha}_3$ & $(1,1,1)$
& $\overline{\alpha}_3$ & $(0,0,0,0)$\\
$\overline{\alpha}_4$ & $\left(\frac{4}{3},\frac{4}{3},\frac{4}{3}\right)$
& $\overline{\alpha}_4$ & $(1,1,1,1)$\\
$\overline{\alpha}_5$ & $\left(\frac{9+\sqrt{21}}{10},\frac{4+\sqrt{21}}{5},\frac{4+\sqrt{21}}{5}\right)$
& $\overline{\alpha}_5$ & $\left(\frac{5}{4},\frac{5}{4},\frac{5}{4},\frac{5}{4}\right)$\\
$\overline{\alpha}_6$ & $\left(\frac{6+\sqrt{6}}{6},\frac{3+\sqrt{6}}{3},\frac{2+\sqrt{6}}{2}\right)$
& $\overline{\alpha}_6$ & $\left(\frac{6+2\sqrt{2}}{7},\frac{5+4\sqrt{2}}{7},\frac{5+4\sqrt{2}}{7},\frac{5+4\sqrt{2}}{7}\right)$\\
\end{tabular}
\caption{Values of $\overline{\alpha}_n$ for $0 \leq n\leq 6$ and $2\leq m \leq 5$.\label{table examples alpha n}}
\end{table}
Notice that $\overline{\alpha}_n$ exists for any $n \in \mathbb{N}$
since $(0,0,...,0)$ is a solution for any $n \in \mathbb{N}$
by the definition of the $p_n$'s.
The proof of the following theorem is similar
to those of~\eqref{eqn pn} and Theorem~\ref{theorem strategy with n cuts}.
\begin{theorem}\
\label{theorem fundamental m rays}
\begin{enumerate}
\item\label{theorem fundamental m rays item solution for ais}
For all $n\in\mathbb{N}$,
the values $a_i$ ($0\leq i < n$) that define $f_n$
satisfy the following properties.
\begin{enumerate}
\item\label{theorem fundamental m rays item ai}
$a_i=p_i(\overline{a})$.

\item\label{theorem fundamental m rays item a0}
$\overline{a}$ is a solution to the system of equations
\[ p_n(\overline{x}) = \rho , \quad
p_{n+1}(\overline{x}) = \rho , \quad
\ldots, \quad
p_{n+(m-2)}(\overline{x}) = \rho .
\]

\item\label{theorem fundamental m rays item cost}
$CR(f_n) = 1+2|\overline{a}|$.
\end{enumerate}

\item The strategy $f_0$ is optimal if and only if $1 \leq \rho \leq \frac{m}{m-1}$.
In this case,
$CR(f_0) = 2(m-1)\rho+1$.

\item For all $n \in \mathbb{N}$,
$$p_{n+m-1}(x) = p_n(\overline{x})\sum_{i=0}^{m-2}x_i - \sum_{i=0}^{n+m-2}p_i(\overline{x}) .$$

\item For all $n\in\mathbb{N}$,
$p_n\left(f_{\infty}(0),f_{\infty}(1),...,f_{\infty}(m-2)\right) = f_{\infty}(n)$.

\item\label{theorem fundamental m rays item alpha n <= m-2}
For all $0 \leq n \leq m-2$,
$\overline{\alpha}_n = (0,0,...,0)$.
Moreover,
$\overline{\alpha}_{m-1} = (1,1,...,1)$
and
$$\overline{\alpha}_m = \left(\frac{m}{m-1},\frac{m}{m-1},...,\frac{m}{m-1}\right) .$$
\end{enumerate}
\end{theorem}

\section{Conclusion}

We have generalized many of our results for searching on a line
to the problem of searching on $m$ rays for any $m \geq 2$.
Even though we could not extend the analysis of the polynomials $p_n$ 
as far as was possible for the case where $m=2$,
we believe this to be a promising direction for future research.
By approaching the problem directly
instead of studying the inverse problem (maximal reach),
we were able to provide exact characterizations of $f_{opt}$ and $CR(f_{opt})$.
Moreover, the sequence of implications
in the proofs of Section~\ref{section searching bounded line}
all depend on~\eqref{factorization pn}, where~\eqref{factorization pn} is
an exact general expression for all roots of all equations $p_n$.
As some readers may have observed,
exact values of the roots of the equation $p_n$ are not required to prove
the results in Section~\ref{section searching bounded line};
we need disjoint and sufficiently tight lower and upper bounds
on each of the roots of $p_n$. In the case where $m > 2$,
finding a factorization similar to~\eqref{factorization pn} appears highly unlikely.
We believe, however, that establishing good bounds 
for each of the roots of the $p_n$ should be possible.
Equipped with such bounds,
the general problem could be solved exactly on $m > 2$ concurrent rays.
We conclude with the following conjecture.
It states that the strategy $f_n$ is uniquely defined for each $n$,
it gives a criterion for the optimal $n$ in terms of $\rho$ (and $m$)
and gives the limit of $f_n$ when $\Dmax \rightarrow \infty$.
\begin{conjecture}\
\begin{enumerate}
\item For all $n\in\mathbb{N}$,
the system of equations of Theorem~\ref{theorem fundamental m rays}-\ref{theorem fundamental m rays item a0}
has a unique solution $\overline{a}^*=(a_0^*,a_1^*,...,a_{m-2}^*)$
satisfying~\eqref{constraint alpha n i}
and such that $|\overline{a}^*| > |\overline{\alpha}_n|$.
Moreover,
there is a unique choice of $\overline{a}$ for $f_n$
and this choice is $\overline{a}=\overline{a}^*$.

\item For a given $\rho$,
if $p_n(\overline{\alpha}_{n+m-1}) \leq \rho < p_n(\overline{\alpha}_{n+m})$,
then $f_n$ is the best strategy
and $|\overline{\alpha}_{n+m-1}| \leq |\overline{a}| < |\overline{\alpha}_{n+m}|$.

\item When $\Dmax \rightarrow \infty$,
then the optimal strategy tends toward $f_{\infty}$.

\item For all $n \in \mathbb{N}$,
$$0 \leq |\overline{\alpha}_n| \leq |\overline{\alpha}_{n+1}| < \frac{m^m}{(m-1)^{m-1}}$$
with equality if and only if $0 \leq n \leq m-3$.
\end{enumerate}
\end{conjecture}

\end{document}